\documentclass{elsarticle}

\usepackage[warn]{mathtext}
\usepackage[T2A]{fontenc}
\usepackage[koi8-r]{inputenc}
\usepackage[english]{babel}

\usepackage[unicode,dvips,pdfpagemode=None,pdfstartview=FitH]{hyperref}
\usepackage{indentfirst}
\usepackage{graphicx}
\usepackage{amssymb}
\usepackage{amsmath}

\newcommand{\SKIP}[1]{}

\begin{document}
\begin{frontmatter}
\title{Search for $\mathbf{\tau}\to\mu+\gamma$ decay at Super $\mathbf{ c -\tau}$ factory.}
\author[Lab3,NSU]{A.V.Bobrov\corref{cor2}}
\author[Lab3,NSU]{A.E.Bondar}
\cortext[cor2]{Corresponding Author}
\address[Lab3]{ Budker Institute of Nuclear Physics RAS,}
\address[NSU]{Novosibirsk State University; Novosibirsk 630090, Russia }


\begin{abstract}

A Monte Carlo study of possible background processes in a search 
for $\tau \to \mu \gamma$ decay has been
performed for conditions of the Super $c-\tau$ factory (CTF)
(at a center-of-mass energy  3.686 GeV, 3.77 GeV and 4.17 GeV).
The background from $\tau^{+}\tau^{-}$ events has been analysed.
Selection criteria for background suppression are suggested
and  necessary requirements on the detector characteristics have been found.
The CTF can successfully
compete with the Super B-factory in a search for $\tau \to \mu \gamma$ decay.

\end{abstract}

\end{frontmatter}

\section{Introduction}

 Decays with Lepton Flavour Violation include mixing leptons from different generations.
They were found in neutrino oscillations.
There is no such evidence in the charged lepton sector.

The decay $\tau\rightarrow\mu+\gamma$ violates lepton flavour.
The Standard Model (SM) prediction for the branching fraction is $\mathcal{B}_{\tau\rightarrow\mu\gamma}<10^{ -50}$  
\cite{3a} (recalculation from neutrino oscillation data).
Extensions of the SM increase  
$\mathcal{B}_{\tau\rightarrow\mu\gamma}$ up to $\sim10^{-8}\div10^{-10}$. 
The best experimental limit is $\mathcal{B}_{\tau\rightarrow\mu\gamma}<4.4\times 10^{-8}$ \cite{4} and
has been obtained at B-factories \cite{11} ($N_{\tau^{+}\tau^{-}}=4.8\times10^{8}$). 

This decay is complicated from the experimental point of view
and interesting for the theory. It is also
convenient to determine detector requirements.

\section{Future experiments for search for $\tau\rightarrow\mu+\gamma$ }

Two types of possible experiments exist to study the $\tau\rightarrow\mu+\gamma$ decay: 
Super-B factory and CTF.

At Super-B factory the total statistics is equal to $N_{\tau^{+}\tau^{-}}\sim6.9\times10^{10}$. 
A feasibility test gives an following upper limit of $2.4\times10^{-9}$ \cite{13}.
 It scales as  $N_{\tau^{+}\tau^{-}}^{-1/2}$
due to unavoidable background from the ISR process ($e^{+}e^{-}\to\tau^{+}\tau^{-}\gamma$). 

CTF \cite{0} is a high luminosity $e^{+}e^{-}$ collider
with  $\mathbf{\cal L}=10^{35}$ cm$^{-2}$s$^{-1}$ and 
 the center-of-mass energy  \SKIP{$\sqrt{s}$} from 2 to 5 GeV. The integrated luminosity is
 10 ab$^{-1}$ after 10-year operation (about $2.5 \times 10^{10}$ $\tau^{+}\tau^{-}$ pairs).
 The CTF detector is a general-purpose $4\pi$ detector. The main points of the physical program are the following:
 studies of charmonium states, spectroscopy of states of light quarks,
physics of $D$ mesons, physics of charmed baryons, $\tau$ lepton physics,
a measurement of the cross section of~  $e^{+}e^{-}\to$ hadrons,
two-photon physics. Background from the ISR process ($e^{+}e^{-}\to\tau^{+}\tau^{-}\gamma$)
 vanishes at low enegry (below $\sim$ 4 GeV). Due to this, the total sensitivity
can be higher at the CTF than at the Super-B, in spite of lower statistics.


\SKIP{

\begin{figure}[ht!]
\begin{center}
\includegraphics[width=0.45\textwidth,clip]{scale.eps}
\caption{Branching scales from statistics}  
\label{f:9}
\end{center}
\end{figure}
  
      }


\section{Theoretical description of $\tau\to\mu+\gamma$ decay}

\SKIP{
\begin{picture}(75,140)
\put(70,0){
\includegraphics[width=105mm,height=45mm,angle=0]{diag.eps}
}

\put(142,113){\makebox{\Large$\nu_{\tau}$}}
\put(283,113){\makebox{\Large$\nu_{\mu}$}}
\put(330,60){\makebox{\Large$\mu^{-}$}}
\put(290,4){\makebox{\Large$\gamma$}}
\put(90,60){\makebox{\Large$\tau^{-}$}}
\put(143,18){\makebox{\Large$W^{-}$}}
\end{picture}

}

The most general theoretical assumptions, namely;
 Lorentz and gauge invariance give us the effective function
 of Lagrangian \cite{5}:
 
${\cal L}_{int} = 
        -\frac{4G_F}{\sqrt{2}}(  
        {m_{\tau }}{A_R}\overline{\tau_{R}}
        {{\sigma }^{\alpha \beta}{\mu_L}{F_{\alpha \beta}}}
       + {m_{\tau }}{A_L}\overline{\tau_{L}}
        {{\sigma }^{\alpha \beta}{\mu_R}{F_{\alpha \beta}}} 
       +  h.c. )
$

The angular distribution in the $\tau$ rest frame:

$\frac{\displaystyle dB}{\displaystyle d\cos\theta}
             = \frac{ \displaystyle B}{\displaystyle 2}\{1+A_{\tau \rightarrow \mu \gamma}P\cos\theta\}$,
             
   $  A_{\tau \rightarrow \mu\gamma} = \frac{|A_L|^2-|A_R|^2}{|A_L|^2+|A_R|^2} ~ B_{\tau \rightarrow \mu\gamma} \propto |A_L|^2+|A_R|^2$.

The longitudinal polarization of the electron beam 
allows a measurement of $A_{\tau \rightarrow \mu\gamma}$ to be performed even
in case of small statistics. It is also possible for $\tau \rightarrow e\gamma$.
A longitudinal polarization gives us new opportunities for background suppression 
(the background from $D\bar{D}$ doesn't have asymmetry).

\section{Background sources and detector requirements}

The most general background sources the following:
 $\tau^{+}\tau^{-}$ decays, 
 QED processes ($e^{+}e^{-}\to\mu^{+}\mu^{-}\gamma\gamma,e^{+}e^{-}\to\mu^{+}\mu^{-}e^{+}e^{-}\gamma$),
 hadron continuum, ISR processes (besides $e^{+}e^{-}\to\tau^{+}\tau^{-}\gamma$),
 pile-up events, accelerator background, $D^{(\ast)}\bar{D}^{(\ast)}$-mesons decays,
 $\psi(2S)$ resonance decays. The last two sources strongly depend on
 the center-of-mass energy  and can be reduced by selection of energy.

Here are the main detector requirements. For the calorimeter, they are
high energy resolution in the range 0.5-1.6 GeV, good time resolution for photons (better than 6 ns),
$\pi^{0}$ reconstruction  efficiency when one photon is extremely soft (up to 3 MeV) and
high photon reconstruction in the $4\pi$ solid angle.
For identification and track system requirements are the following: high  quality of $\pi/\mu$ separation in the momentum range 0.5-1.6 GeV/c,
and high reconstruction efficiency for charged particles in the $4\pi$ solid angle.

\section{Simulation}

We used fast simulation taking into account energy
and coordinate resolution of the calorimeter,
a threshold of photon detection
 (20 MeV) and  momentum resolution of charged particles.
 Particle identification wasn't simulated. Monte Carlo search for
1.5\% and 2.5 \% energy resolution of the calorimeter was performed. 

The  TAUOLA \cite{3} generator was used for $\tau$ lepton
decay simulation.

\SKIP{

\begin{center}
Detector simulation
\end{center}
\begin{itemize}
\item{fast simulation}
\item{energy resolution (response function is match of an exponent with a Gaussian)}
\item{coordinate resolution of calorimeter - 1 cm}
\item{acceptance \  $20^{\circ}<\Theta<160^{\circ}$}
\item{threshold of photons registration  20 MeV}
\item{momentum resolution }
\item{efficiency 100\% }
\item{particle identification is not simulated}
\item{energy resolution  is 1.5\% and 2.5\% }

\end{itemize}

\begin{center}
Decays simulation
\end{center}

  For simulation of $\tau$ decays the TAUOLA generator was used.
  The formfactor of $\tau^{-}\rightarrow\pi^{-}+\pi^{0}+\nu_{\tau}$ decay 
  has been  modified.

  For  $\tau\rightarrow\mu+\gamma$ generation  $\tau\rightarrow\pi+\nu_{\tau}$
  was used  with   $m_{\pi}\rightarrow m_{\mu}$. 
  If $|A_{\tau \rightarrow \mu\gamma}|=1$,
  the angular distributions is the same.

      }

\section{$\mathbf{\tau^{+}\tau}^{-}$ background}

Two types of background exist: direct and combinatorial.
Direct background is caused by the decay products of 
one tau-lepton  imitating a signal, e. g., 
$\tau^{-}\rightarrow\mu^{-}+\gamma+\nu_{\tau}+\bar{\nu}_{\mu}$ and
$\tau^{-}\rightarrow\pi^{-}+\pi^{0}+\nu_{\tau }$.
The combinatorial background is a process when a muon (or pion) from decay
of one tau and a photon from decay of another tau imitate a
signal, e. g., is $\tau^{+}+\tau^{-}\to\mu\bar{\nu}_{\mu}\nu_{\tau}+
\pi^{+}\pi^{0}\bar{\nu}_{\tau}\to\mu\gamma+\pi^{+}\gamma_{soft}\bar{\nu}_{\mu}\nu_{\tau}\bar{\nu}_{\tau}$.
This type of background depends on the decay mode of the second (non-signal) tau.

The following parameters for event selection were used:  
$\Delta E_{+}>\Delta E>\Delta E_{-}$, $\Delta m_{bc}^{+}>m_{bc}-m_{\tau}>\Delta m_{bc}^{-}$,
the square of neutrino mass (for semileptonic decay of a non-signal $\tau$)  $m_{rec+}^{2}>m_{rec}^{2}>m_{rec-}^{2}$,
the mass of $\mu\nu$ (or $\mu\nu\bar{\nu}$ for
leptonic decays of the second tau) system, $\pi^{0}$ veto.

\section{Results and conclusions }


Number of events ($S$-signal, $B_{\pi}$-background from $\pi$, $B_{\mu}$-background from $\mu$).
The total statistics are $N_{\tau^{+}\tau^{-}}=3.2\times10^{10}$  and the branching fraction  is $10^{-9}$.

      \begin{table}[h!]
\small
      \begin{tabular}{|l|l|l|l|l|l|l|}
                          \hline            
                          
      $\sqrt{s}$ GeV &  3.686   &  3.686   & 3.77   & 3.77   &4.17   & 4.17   \\      
      $\frac{\displaystyle\sigma_{E}}{\displaystyle E}$  &1.5\%  & 2.5\%& 1.5\% & 2.5\%& 1.5\% & 2.5\% \\

          \hline            
           \multicolumn{1}{|r|}{$ \pi^{-}\nu_{\tau}$ \  $S$}   & 4.3      & 4.0         &  4.3       &  4.0      & 3.0  & 2.7   \\
                        \multicolumn{1}{|r|}{$B_{\mu}$}  & $0.9^{+1.2}_{-0.6}$      & $5.3^{+2.0}_{-1.5}$          & $4^{+1.8}_{-1.3}$        &  $4.7_{-1.7}^{+2.5}$      & $1.4^{+3.3}_{-1.1}$  &   $7.1^{+4.9}_{-3.1}$         \\
            \multicolumn{1}{|r|}{$B_{\pi}$}       & $14.6^{+3.2}_{-2.2}$      &  $58^{+5}_{-4}$        &$54\pm6$         &  $164\pm10$      & $15.6^{+2.8}_{-1.8}$  &    $62^{+2.7}_{-1.8}$        \\                        
          \hline                                  
           \multicolumn{1}{|r|}{$ \pi^{-}\pi^{0}\nu_{\tau}$ \  $S$}    & 9.3      &   6.7       & 8.7        &  6.7      & 5.7  &   5.3         \\
                        \multicolumn{1}{|r|}{$B_{\mu}$}          &  $5.8^{+2.7}_{-2.0}$   &  $8.9^{+3.0}_{-2.3}$        &   $8.1^{+2.7}_{-2.0}$      &  $10^{+3.0}_{-2.5}$      & $10.6^{+3.3}_{-2.6}$  &  $9.1^{+5.4}_{-3.7}$          \\
            \multicolumn{1}{|r|}{$B_{\pi}$}          & $21^{+2.2}_{-1.7}$      &   $100^{+3.8}_{-3.2}$       &    $26\pm3$      & $127\pm4$       &$32\pm3$   &        $147^{+5}_{-4}$    \\                        
          \hline                               
           \multicolumn{1}{|r|}{$ \mu^{-}\bar{\nu}_{\mu}\nu_{\tau}$ $S$}          & 4.7      &    4.3      &   4.7      & 3.3       &  2.6 & 2.3           \\
                     \multicolumn{1}{|r|}{$B_{\pi}$}          & $20^{+4.3}_{-2.6}$      & $82.0^{+5.5}_{-5.0}$ &  $30^{+6}_{-4}$       &         $127\pm7$      &$70\pm8$  &$ 138\pm7$           \\                        
          \hline                                  
           \multicolumn{1}{|r|}{$ e^{-}\bar{\nu}_{e}\nu_{\tau}$ $S$}          & 4.7      & 4.0      & 4.3      &   3.0     &  2.7 &  2.6          \\
                     \multicolumn{1}{|r|}{$B_{\pi}$}          & $16\pm1$      & $74\pm2.5$         &   $17\pm1$      &    $101\pm7$    & $22\pm8$  &      $108\pm6$      \\                        
          \hline                                  
           \multicolumn{1}{|r|}{$ \pi^{-}\pi^{0}\pi^{0}\nu_{\tau}$ $S$}          &   2.6    &  2.5        &  2.6       &  2.4      & 2.6  &   2.5         \\
                     \multicolumn{1}{|r|}{$B_{\pi}$}          & $7.5\pm0.5$      &   $35\pm1$       &  $8.0\pm0.5$       & $46\pm3$       & $10.5\pm3.0$  &    $51\pm3$        \\                        
          \hline                                  
           \multicolumn{1}{|r|}{$ \pi^{-}\pi^{+}\pi^{-}\nu_{\tau}$ $S$}          &  2.9     &     2.7     &    2.7     &    3.0    & 3.1  &    2.9        \\
                     \multicolumn{1}{|r|}{$B_{\pi}$}          & $7.5\pm0.5$      &  $35\pm1$        &  $8.0\pm0.5$       & $46\pm3$       & $10.5\pm3.0$  &     $51\pm3$       \\                        
          \hline

      \end{tabular}
      \end{table}

\normalsize

\begin{figure}[ht!]
\begin{center}
\includegraphics[width=0.4\textwidth,clip]{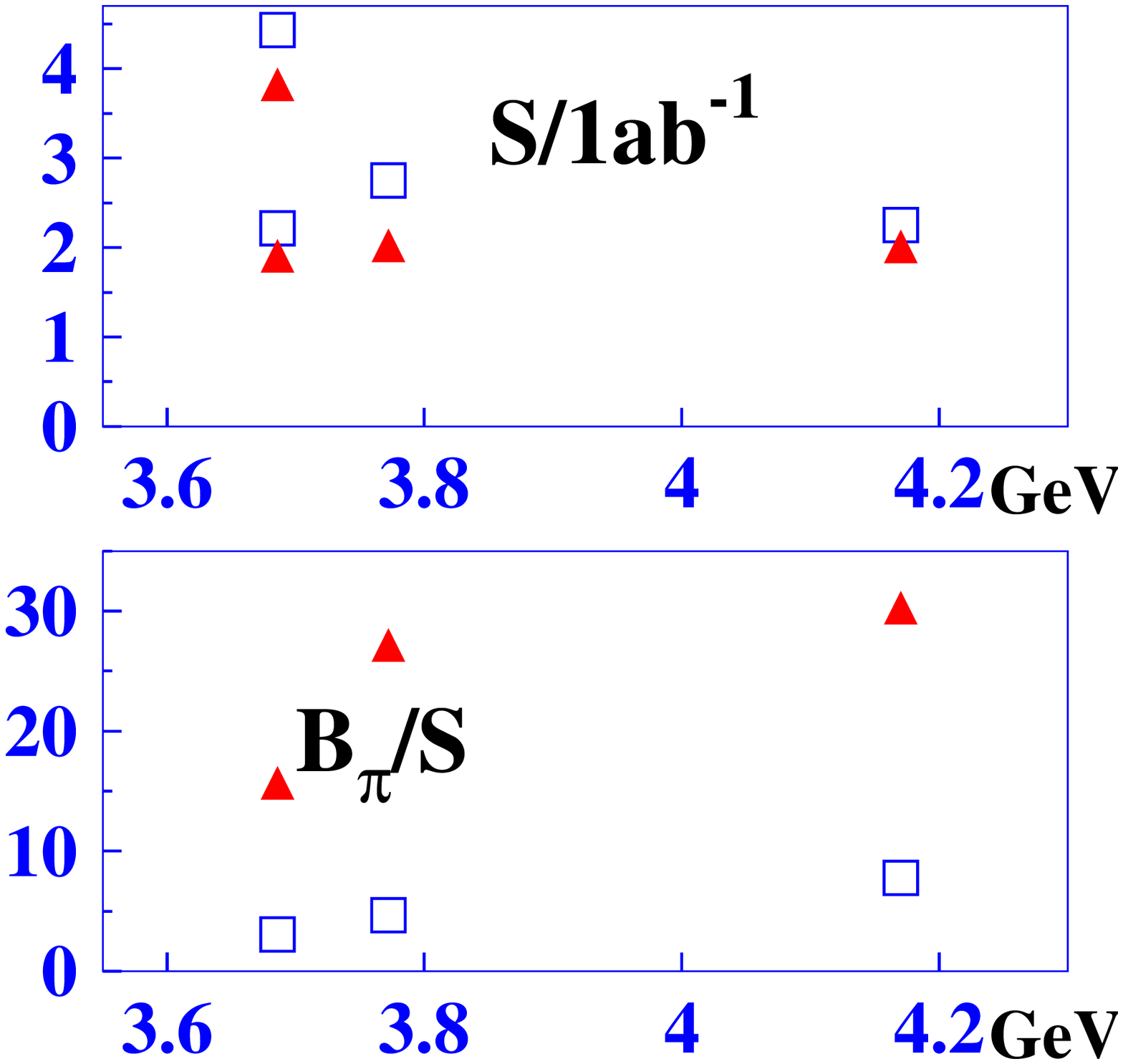}
\caption{Number of signal events (at $\mathcal{B}_{\tau\to\mu\gamma}=10^{-9}$)
for 1 ab$^{-1}$ and background signal ratio for pions, 
squares -- 1.5\% triangles -- 2.5\% resolutions.
}  
\label{f:9}
\end{center}
\end{figure}

The total number of events at 10 ab$^{-1}$ (according to scenario 1.5 ab$^{-1}$ at 3.686 GeV, 3.5 ab$^{-1}$ at 3.77 GeV and 2 ab$^{-1}$ at 4.17 GeV; all points beyond $\tau^{+}\tau^{-}$
threshold)  is the following:
$S=20$ $B_{\pi}=95^{+4.1}_{-3.5}$, $B_{\mu}=8^{+1.40}_{-0.95}$.

So, necessary (but not sufficient) detector requirements for $\tau\to\mu\gamma$ search are  
1 per 40 suppression level $\pi/\mu$, good energy resolution (better than 2\%).
Full reconstrution second tau decay (beside neutrino) it is necessary for background suppression.
The total upper limit on the decay  will be about $10^{-9}$.

\SKIP{
\section{Conclusions}

   \begin{picture}(200,212)
\put(20,50){
\parbox{1\textwidth}{
\begin{center}
  \begin{itemize}

  \item{full reconstruction of the second $\tau$ is necessary for background suppression}

   \item{ratio  $\frac{\displaystyle S}{\displaystyle B}$  increases at 4 times than  energy resolution
   
   varies from 1.5\%  to 2.5\%   and  increases at 1.6 times   
   
   than  energy  ($\sqrt{s}$) varies from
     3.686 GeV up to 3.77 GeV }
     
    \item{ 1/40  $\mu/\pi$ suppression will be enough for
       $\tau^{+}\tau^{-}$ background events}    


   \end{itemize}
\end{center}

}
  }

\put(1100,50){
\parbox{0.49\textwidth}{
\begin{center}
  \begin{itemize}

    \item{decay $\tau \to \mu \gamma$ is convenient for detector optimisation }    

    \item{it is possible to eliminate background from $\tau^{+}\tau^{-}$ up to level $\cal{B}  \sim 10^{-9}$ }

    \item{CTF can compete with Super-B in a search for $\tau \to \mu \gamma$ }        



   \end{itemize}
\end{center}

}

}  
   \end{picture}


      }

\end{document}